\documentclass{article}%
\usepackage{amssymb}
\usepackage{amsmath}
\usepackage{amsfonts}
\usepackage{graphicx}%
\begin{document}
\[
\]
{\Large Some remarks about extremum principles involving the rate of entropy
production}

M. Salis$^{a}$

\textit{Dipartimento di Fisica - Universit\`{a} di Cagliari- Cittadella
Universitaria , 09042 Monserrato-Cagliari, Italy}%
\[
\]
The minimum rate of entropy production (MREP) and the least dissipation energy
(LDE) principles are re-examined concerning continuous systems in stationary
nonequilibrium states. By means of simple considerations on coefficients of
phenomenological laws and by taking into account balance equations, an
Onsager-like potential density for the stationary state is obtained. In the
case of a single state variable it is found that the potential density
describes a saddle-shaped surface around the stationary point in the domain of
the force (state variable) and the flux. The saddle point may be considered a
maximum along a LDE path (fixed force) and a minimum along a MREP path (fixed
phenomenological coefficient). The latter allow for the Glansdorff-Prigogine
local potential.%

\[
\]
\textbf{I.} \textbf{INTRODUCTION}

Thermodynamics establishes that, under certain conditions, systems in
stationary nonequilibrium states (SNES) have a minimum rate of entropy
production (MREP) compatible with system constraints.$^{1}$ The MREP principle
(or criterion) requires that linear phenomenological relations between
generalized fluxes and forces hold. This condition is satisfied when slight
deviations from equilibrium take place uniformly in the whole system
volume.$^{2}$ Beyond this range some additional assumptions are required in
order to apply a suitably modified principle.$^{3}$ The matter is not so well
defined in continuous systems when deviations from equilibrium are not
uniformly distributed. Remaining near equilibrium, several problems can be
assessed successfully provided the flux laws are suitably linearized.$^{4}$
But in most case the linearization procedure presents some difficulties.
Actually, according to statistical mechanics, linear laws hold when local
thermodynamic equilibrium (LTE) conditions are satisfied.$^{5}$ But, in this
case linearity is meant in the sense of deviations from the local equilibrium.
Thus, in general, phenomenological laws are not strictly linear. The matter is
further complicated by a certain freedom in the choice of forces and
conjugated fluxes.$^{6}$ Thus, for example, in the case of heat conduction the
MREP principle apparently works$^{5}$ or fails$^{7}$ according to the used conventions.

Glansdorff and Prigogine$^{1,8}$ formulated a general evolution criterion for
irreversible processes which is independent of the actual form of the
phenomenological coefficients. Based on this criterion, they devised a
procedure to define a (local) potential for the SNES which coincides (unless a
numeric factor) with the entropy production functional in the strictly linear
range. Beyond this range, the local potential allows for a (generalized) MREP
principle provided certain restrictions to the entropy production functional
are assumed.$^{9}$

To give a thermodynamic framework to phenomenological laws, Onsager and
Machlup proved a least dissipation energy (LDE) principle for near-equilibrium
processes.$^{10,11}$ In the assumption of Gaussian behaviour of extensive
state variables, the nonequilibrium processes are Markovian. A system
displaced from equilibrium by a fluctuation decays back to equilibrium by
following, on the average, the same empirical laws for the decay of the system
displaced by an external constraint.$^{11}$ According to this theory, the
probability associated with a given decay path is to be searched for by means
of an extremum principle of a suitable functional. The most probable path
(with fixed forces) is the one established by phenomenological laws, i. e., by
the LDE principle.

The theory developed by Onsager and Machlup has given insights for the
formulation of several Hamilton-based variational principles in the field of
hydrodynamics. We do not give accounts of the many works devoted to this topic
(see for example ref. 9). In this paper we are interested in the possible
connection between the Onsager (LDE) and the Glansdorff-Prigogine (extended
MREP) principles for continuous systems. Specifically, we ask for a
variational criterion including both the two principles to give a unified and
concise representation of our knowledge about the SNES. To achieve its
formulation we will pursue a simple task: the criterion must account for the
(entropy and mass) balance equations regardless of the actual form of
phenomenological coefficients.

To deal with this matter the paper is planned as follows. In Section 2 we will
examine some features concerning the rate of entropy production as established
by thermodynamics. In section 3 we will derive the variational criterion for
the SNES. The extremum properties of a functional dependent on forces (state
variables) and fluxes will be investigated. In the simplest case of a single
(state) variable the density of the functional describes a saddle-shaped
surface in the domain of force and flux. The extended MREP and LDE principles
will be apparent by restrictions of the potential to special deviation paths.
In the field of thermodynamics, restriction of the functional to the MREP path
allows for the Glansdorff-Prigogine local potential.
\[
\]
\textbf{II} \textbf{Entropy production}

The definition of entropy production follows from the thermodynamic
construction of the local entropy balance equation$^{2,5}$%
\begin{equation}
\sigma_{S}=\frac{\partial\rho s}{\partial t}+\nabla\cdot\mathbf{J}_{STOT}
\label{MREP2}%
\end{equation}
where $s$ stands for the specific entropy , $\rho$ for mass density ,
$\mathbf{J}_{STOT}$\ for total entropy flux. The time derivative in the local
frame is connected to the barycentre derivative by the relation%

\begin{equation}
\frac{\partial\rho s}{\partial t}=\rho\frac{ds}{dt}-\nabla\cdot\rho
s\mathbf{v} \label{MREPXX2}%
\end{equation}
where $\mathbf{v}$ stands for the centre of mass velocity. By inserting eq.
(\ref{MREPXX2}) in eq. (\ref{MREP2}) we define entropy flux%

\begin{equation}
\mathbf{J}_{S}=\mathbf{J}_{STOT}-\rho s\mathbf{v}~. \label{MREPXX3}%
\end{equation}
The starting point of the construction procedure is the Gibbs equation
(assumed to hold in LTE conditions) in which energy and mass balance equations
have been inserted. These latter read\
\begin{equation}
\frac{\partial\rho_{i}}{\partial t}+\nabla\cdot\left(  \mathbf{j}_{i}+\rho
_{i}\mathbf{v}\right)  =\sum\nu_{ij}\xi_{j} \label{MREP3}%
\end{equation}
\ where $\rho_{i}$ stands for the density of the i-th system component,
$\mathbf{j}_{i}$ for diffusion flow, $\nu_{ij}\xi_{j}$ is the rate of mass
production of the i-th component in the j-th chemical reaction with reaction
rate $\xi_{j}$ and stoichiometric coefficient $\nu_{ij}$ . The entropy flux
can be written in general as%

\begin{equation}
\mathbf{J}_{S}=\left(  \mathbf{J}_{Q}-\sum_{i}\varphi_{i}~\mathbf{j}%
_{i}\right)  /T \label{MREPYY4}%
\end{equation}
where $\mathbf{J}_{Q}$ stands for the heat flux and $\varphi_{i}$ for
electrochemical potential of the i-th component$^{13}$. We are interested in
those states where the barycentre derivatives vanish. Moreover, we consider
the case of mechanical equilibrium where derivatives of $\mathbf{v}$ vanish
everywhere in the system$^{6}$. In these conditions, the density of entropy
production is$^{12}$%

\begin{equation}
\sigma_{S}=-\sum_{i}\mathbf{j}_{i}\cdot\nabla\frac{\varphi_{i}}{T}%
+\mathbf{J}_{Q}\cdot\nabla\frac{1}{T}-\sum_{i}\frac{\xi_{i}\Gamma_{i}}{T}
\label{MREP1}%
\end{equation}
where $\Gamma_{i}=\sum_{j}\nu_{ij}\varphi_{j}$ stands for the affinity of the
set of reactions involved in the production of the i-th component. When the
centre of mass motion is determined by flows $\mathbf{j}_{i}$ we have the
condition $\sum\mathbf{j}_{i}=0$. In this case the entropy production should
be re-written in order to present only independent flows$^{12}$. Equation
(\ref{MREP1}) can be used, for example, to investigate thermoelectric
processes in semiconductors. In this case, chemical reactions are replaced by
electron and hole excitation processes.

Formally, $\sigma_{S}$ can be written as a product of generalized fluxes
$\mathbf{J}$ and forces $\mathbf{X}$%

\begin{equation}
\sigma_{S}=\sum_{k}\mathbf{J}_{k}\cdot\mathbf{X}_{k}~. \label{MREP4}%
\end{equation}
Fluxes are assumed to be functions of the forces (phenomenological laws):%

\begin{equation}
\mathbf{J}_{k}=\sum_{l}L_{kl}\mathbf{X}_{l}~. \label{MREP5}%
\end{equation}
By inversion of eqs. (\ref{MREP5}) we have%

\begin{equation}
\mathbf{X}_{k}=\sum_{l}M_{kl}\mathbf{J}_{l} \label{MREP5BIS}%
\end{equation}
where $M=L^{-1}$. According to Curie's theorem, phenomenological coefficients
are scalars in isotropic systems so that only fluxes and forces of the same
vectorial (or tensorial) character are related. The requirement of microscopic
time reversibility of processes near equilibrium allow for symmetry of
coefficients associated with cross effects (this does not hold in cases of
magnetic and Coriolis forces$^{6,12,14}$) :%

\begin{equation}
L_{ik}=L_{ki} \label{MREP6}%
\end{equation}
Owing to the detailed balance not all the possible definitions of forces and
conjugated fluxes are allowed$^{6}$ ( two important conventions consider
"energy transport per molecule" or the "entropy transport per molecule"$^{15}%
$). In this connection, we point out that a variational principle correctly
formulated must not depend on the convention adopted.

The entropy production of the whole system is (integrated over the system volume)%

\begin{equation}
P=%
%TCIMACRO{\dint }%
%BeginExpansion
{\displaystyle\int}
%EndExpansion
\sum_{k}\mathbf{J}_{k}\cdot\mathbf{X}_{k}d^{3}x \label{MREP7}%
\end{equation}
By considering the steady state as a reference state ($\mathbf{J}%
_{k}=\mathbf{J}_{0k}+\delta\mathbf{J}_{k},\mathbf{X}_{k}=\mathbf{X}%
_{0k}+\delta\mathbf{X}_{k},$), the first variation of P at the point
$(X_{0},J_{0})$ due to small deviations of forces is (constant Onsager's coefficients)%

\begin{equation}
\delta^{(1)}P=%
%TCIMACRO{\dint }%
%BeginExpansion
{\displaystyle\int}
%EndExpansion
\sum_{k}\mathbf{X}_{0k}\cdot\delta\mathbf{J}_{k}d^{3}x+%
%TCIMACRO{\dint }%
%BeginExpansion
{\displaystyle\int}
%EndExpansion
\sum_{k}\mathbf{J}_{0k}\cdot\delta\mathbf{X}_{k}d^{3}x \label{MREP8}%
\end{equation}
The second variation is%

\begin{equation}
\delta^{(2)}P=%
%TCIMACRO{\dint }%
%BeginExpansion
{\displaystyle\int}
%EndExpansion
\sum_{k}\delta\mathbf{J}_{k}\cdot\delta\mathbf{X}_{k}d^{3}x \label{MREP9}%
\end{equation}
Thus $\Delta P=\delta^{(1)}P+\delta^{(2)}P$ ; no higher terms appear with the
settings dealt with (see next section). Equation (\ref{MREP9}) defines the
excess of entropy production.$^{1}$ If we consider equilibrium as the
reference state ($X_{0k}=0$;$J_{0k}=0$), we remain with $\Delta P=P=%
%TCIMACRO{\dint }%
%BeginExpansion
{\displaystyle\int}
%EndExpansion
\sum_{k}\mathbf{J}_{k}\cdot\mathbf{X}_{k}d^{3}x$ so that $\Delta P\geq0$ and
$d\Delta P/dt\leq0$ ($=0$ at the stationary state) which explains the MREP principle.

Let us we consider a simple example to clarify some points presented in the
introduction. The diffusion under uniform temperature (for simplicity's sake
we do not consider the Doufour and Soret effects) in absence of sources shows
$\sigma_{S}=-j~\nabla\rho/\rho^{n}$ (we leave unspecified exponent n). A
possible choice of the force and the conjugated flux is%

\begin{equation}
\mathbf{X}=-\frac{1}{\rho^{n}}\nabla\rho\ \ ~\ \ \ \ \mathbf{J}=\mathbf{j}%
=-\frac{L}{\rho^{n}}\nabla\rho\label{MREP10}%
\end{equation}
The variational calculus is applied to the functional $P[\rho]=\int
X(\rho)J(\rho)d^{3}x$ \ by keeping fixed the variable at the system boundary:%
\[
\delta P\left(  \rho\right)  =\delta\int\frac{L}{\rho^{2n}}\left(  \nabla
\rho\right)  ^{2}d^{3}x=\int\left[  \left(  \frac{\partial}{\partial\rho
}-\nabla\frac{\partial}{\partial\nabla\rho}\right)  \frac{L}{\rho^{2n}}\left(
\nabla\rho\right)  ^{2}\right]  \delta\rho d^{3}x=0
\]
that is,%

\begin{equation}
\int\left\{  \left[  \frac{d}{d\varrho}\left(  \frac{L}{\varrho^{2n}}\right)
+2n\frac{L}{\varrho^{2n+1}}\right]  \left(  \nabla\rho\right)  ^{2}-\frac
{2}{\rho^{n}}\nabla\left(  \frac{L}{\rho^{n}}\nabla\rho\right)  \right\}
\delta\rho~d^{3}x=0 \label{MREP11}%
\end{equation}
Equation (\ref{MREP11}) is apparently written in a complicated form, but it
clearly shows that to account for the right transport equation ($L\nabla
\rho/\rho^{n}=const$) we require that $dL/d\rho=0$. The latter condition
appears excessively restrictive since in some important cases (Fourier's law ,
Fick's law) $L\varpropto\rho^{n}$ . A better choice may be:
\begin{equation}
\mathbf{X}=-\nabla\rho\ \ ~\ \ \ \ \mathbf{J}=\mathbf{j}/\rho^{n}=-L^{\prime
}\nabla\rho\label{MREP11XBIS}%
\end{equation}
where $L^{\prime}$ is a suitable constant coefficient. However, this choice
does not match the transport features if the actual diffusion process requires
$L^{\prime}$ to be dependent on variable $\rho$.%
\[
\]
\textbf{III.} \textbf{STATIONARY STATES BEYOND SMALL DEVIATIONS FROM
EQUILIBRIUM.}

Henceforth, we will take SNES as the reference state from which deviations of
variables are considered. To represent deviations of fluxes one can use
expansions with respect to forces.$^{1}$ But in this way information about
phenomenological coefficients are lost. In a more conservative representation
deviations of coefficients are explicitly considered. Thus, by taking into
account of expansions%

\begin{equation}
\mathbf{X}_{j}\left(  \rho\right)  =\mathbf{X}_{0j}+\delta^{(1)}\mathbf{X}%
_{j}\left(  \rho\right)  +\delta^{(2)}\mathbf{X}_{j}\left(  \rho\right)  +...
\label{MREPXYZ12}%
\end{equation}
and%

\begin{equation}
L_{ij}\left(  \rho\right)  =L_{0ij}+\delta L_{ij}\left(  \rho\right)
+\delta^{(2)}L_{ij}\left(  \rho\right)  +... \label{MREPXZZ12}%
\end{equation}
we obtain from eq. (\ref{MREP5}):%
\begin{equation}
\delta^{(1)}\mathbf{J}_{i}\left(  \rho\right)  =\sum L_{0ij}\delta
^{(1)}\mathbf{X}_{j}\left(  \rho\right)  +\sum~\mathbf{X}_{0j}^{(1)}~\delta
L_{ij}\left(  \rho\right)  ~. \label{MREP12}%
\end{equation}
For convenience, in eqs (\ref{MREPXYZ12}) and (\ref{MREPXZZ12}) we have
indicated the functional dependence on state variables (for brevity symbolized
by $\rho$) of all terms except the quantities $X_{0j}$ and $L_{0ij}$ which
pertain to the reference state. By taking into account that$^{1}$
\begin{equation}%
%TCIMACRO{\dint }%
%BeginExpansion
{\displaystyle\int}
%EndExpansion
\sum_{k}\mathbf{J}_{0k}\cdot\delta^{(1)}\mathbf{X}_{k}\left(  \rho\right)
d^{3}x=0 \label{MREPXX13}%
\end{equation}
eqs (\ref{MREP12}) and (\ref{MREP8}) provide a way to represent SNES whatever
conventions are used. Based on eq. (\ref{MREPXX13}) we may write%

\begin{equation}
\delta^{(1)}P[\rho]=%
%TCIMACRO{\dint }%
%BeginExpansion
{\displaystyle\int}
%EndExpansion
\sum_{k}X_{0k}\delta^{(1)}J_{k}\left(  \rho\right)  d^{3}x \label{MREPXAgg1}%
\end{equation}
which represents the variation of entropy production at the point $\rho_{0}$.
In general, variation $\delta^{(1)}P[\rho]$ is different from 0 at the
stationary state. To search for a potential showing vanishing variation at
this state it is convenient to use the transformation

\begin{equation}%
%TCIMACRO{\dint }%
%BeginExpansion
{\displaystyle\int}
%EndExpansion
\sum_{k}\mathbf{X}_{0k}\cdot\delta^{(1)}\mathbf{J}_{k}\left(  \rho\right)
d^{3}x=\delta^{(1)}\frac{1}{2}%
%TCIMACRO{\dint }%
%BeginExpansion
{\displaystyle\int}
%EndExpansion
\sum_{kl}M_{0kl}\mathbf{J}_{k}\left(  \rho\right)  \cdot\mathbf{J}_{l}\left(
\rho\right)  d^{3}x \label{MREP14BIS}%
\end{equation}
where $M_{0}=L_{0}^{-1}$. Thus, by taking into account of eqs (\ref{MREP8}),
(\ref{MREPXX13}) and (\ref{MREP14BIS}) we obtain%

\begin{equation}
\delta^{(1)}\left\{  P[\rho]-\frac{1}{2}%
%TCIMACRO{\dint }%
%BeginExpansion
{\displaystyle\int}
%EndExpansion
\sum_{kl}M_{0kl}\mathbf{J}_{k}(\rho)\cdot\mathbf{J}_{l}(\rho)d^{3}x\right\}
=0 \label{MREP15}%
\end{equation}
from which we define the functional%

\begin{equation}
\Psi\lbrack\rho]=%
%TCIMACRO{\dint }%
%BeginExpansion
{\displaystyle\int}
%EndExpansion
\sum_{k}\mathbf{J}_{k}\left(  \rho\right)  \cdot\mathbf{X}_{k}\left(
\rho\right)  d^{3}x-\frac{1}{2}%
%TCIMACRO{\dint }%
%BeginExpansion
{\displaystyle\int}
%EndExpansion
\sum_{kl}M_{0kl}\mathbf{J}_{k}(\rho)\cdot\mathbf{J}_{l}(\rho)d^{3}x
\label{MREP15BIS}%
\end{equation}
and the associated density%

\begin{equation}
\psi(\rho)=\sum_{k}\mathbf{J}_{k}(\rho)\cdot\mathbf{X}_{k}(\rho)-\frac{1}%
{2}\sum_{kl}M_{0kl}\mathbf{J}_{k}(\rho)\cdot\mathbf{J}_{l}(\rho)
\label{MREPYY15}%
\end{equation}
According to definition (\ref{MREP15BIS}), we can rewrite eq. (\ref{MREP14BIS}%
) as $\delta^{(1)}\Psi\lbrack\rho]=0$ (at SNES). \ It is worth to point out
that the second deviation $\delta^{(2)}\psi\left(  \rho\right)  $ does not
present terms such as $\sum_{k}\mathbf{X}_{0k}\cdot\delta^{(2)}\mathbf{J}%
_{k}\left(  \rho\right)  $. This can be verified from eq. (\ref{MREPYY15}). by
taking into account eq. (\ref{MREP5BIS}) written for the stationary state
point, that is,
\begin{equation}
\delta^{(2)}\psi(\rho)=\sum_{k}\delta^{(1)}\mathbf{J}_{k}(\rho)\cdot
\delta^{(1)}\mathbf{X}_{k}(\rho)-\frac{1}{2}\sum_{kl}M_{0kl}\delta
^{(1)}\mathbf{J}_{k}(\rho)\cdot\delta^{(1)}\mathbf{J}_{l}(\rho) \label{MREP22}%
\end{equation}
where the sums $\sum_{k}\mathbf{X}_{0k}\cdot\delta^{(2)}\mathbf{J}_{k}(\rho)$
multiplied by opposite signs are canceled out. In eq. (\ref{MREP22}) we have
omitted the sum $\sum_{k}\mathbf{J}_{0k}\cdot\delta^{(2)}\mathbf{X}_{k}(\rho)$
since its integral vanishes as in eq. (\ref{MREPXX13}). On these grounds, we
can define the functional $\Psi\lbrack\rho,J]$ which has the form
(\ref{MREP15BIS}) but in which the function $\mathbf{J}_{k}\left(
\rho\right)  $ is replaced by the variable $\mathbf{J}_{k}$. Analogously, we
follows the same rule to define the density $\psi(\rho,\mathbf{J})$ from eq.
(\ref{MREPYY15}). Thus, the variational equation
\begin{equation}
\delta\Psi=0 \label{MREPZZ15}%
\end{equation}
($\Psi$ be stationary at the SNES) can be solved with respect to the state
variables (see eqs \ref{MREPXYZ12} \ to \ref{MREP12}) or also with respect to
the fluxes. In this case, the associated Eulero-Lagrange equations are%

\begin{equation}
\left(  \partial\psi/\partial\mathbf{J}_{l}\right)  _{SNES}=0~.
\label{MREPZZ16}%
\end{equation}
which go back to the phenomenological laws at the SNES. The set of
eqs\ (\ref{MREPZZ16}) devises a local extremum problem of the function $\psi$
with respect to the fluxes by keeping the forces (state variables) fixed to
their reference state values. To assure the existence of such an extremum it
is sufficient for a general quadratic form $f(\eta)=\sum_{kl}M_{0kl}\eta
_{k}\eta_{l}$\ to have a definite sign. In particular, $\psi\left(
\rho,\mathbf{J}\right)  $ has a maximum (fixed forces) if the quadratic form
is positive definite.

In general, the functional $\Psi$ does not show an extremum at SNES. However,
as seen above, we can search for some special variation paths, crossing the
SNES point in the domain of forces (state variables) and fluxes where this can
be established in a restricted sense. To this end, we extend the problem
(\ref{MREPZZ16}) to any set of fixed forces lying within a small interval
around the SNES point. The locus of maxima thus found satisfies the relation

\begin{equation}
\mathbf{X}_{k}=\sum_{l}M_{0kl}\mathbf{J}_{l} \label{MREPZZBis17}%
\end{equation}
which by inversion gives%

\begin{equation}
\mathbf{J}_{k}=\sum_{l}L_{0kl}\mathbf{X}_{l}~. \label{MREPZZ17}%
\end{equation}
Note that the points represented by eqs \ (\ref{MREPZZ17}) does not satisfy
the LTE conditions which are fitted by eq. (\ref{MREP5}). Thus, deviations of
fluxes defined by%

\begin{equation}
\delta\mathbf{J}_{k}\left(  \rho\right)  =\sum_{l}L_{0kl}\delta\mathbf{X}%
_{l}\left(  \rho\right)  \label{MREPXYZ17}%
\end{equation}
does not introduce any additional condition such as $\delta L_{kl}\left(
\rho\right)  =0$. In the view of a local LDE principle, deviations
(\ref{MREPXYZ17}) may be considered as fluctuations from the (SNES) local
equilibrium. From substitution of eq. (\ref{MREPZZ17}) into eq.
(\ref{MREPYY15}) we obtain%

\begin{equation}
\widetilde{\psi}\left(  \rho\right)  =\frac{1}{2}\sum_{kl}L_{0kl}%
\mathbf{X}_{k}\left(  \rho\right)  \cdot\mathbf{X}_{l}\left(  \rho\right)
\label{MREPZZ18}%
\end{equation}
Accordingly, we define the functional%

\begin{equation}
\widetilde{\Psi}[\rho]=\frac{1}{2}\int\sum_{kl}L_{0kl}\mathbf{X}_{k}\left(
\rho\right)  \cdot\mathbf{X}_{l}\left(  \rho\right)  d^{3}x \label{LDE2}%
\end{equation}
Variations of $\widetilde{\Psi}[\rho]$ imply variations of $\Psi\lbrack
\rho,J]$\ along paths defined by eq. (\ref{MREPXYZ17}). Now, the extremum
property of $\widetilde{\Psi}[\rho]$\ depends on the sign of $\delta
^{(2)}\widetilde{\psi}\left(  \rho\right)  $, that is,%
\begin{equation}
\delta^{(2)}\widetilde{\psi}\left(  \rho\right)  =\frac{1}{2}\sum
L_{0ij}\delta\mathbf{X}_{i}\left(  \rho\right)  \delta\mathbf{X}_{j}\left(
\rho\right)  ~. \label{LDE1}%
\end{equation}
\ The same arguments presented for the problem (\ref{MREPZZ16}) can be used
for eq. (\ref{LDE1}). If $f(\eta)$ is positive definite it follows that
$\delta^{(2)}\widetilde{\psi}\left(  \rho\right)  \geq0$ so that
$\widetilde{\Psi}[\rho]$\ shows a minimum at the SNES. Thus the extremum
properties of the functional $\Psi\lbrack\rho,\mathbf{J}]$\ along the two
paths dealt with are determined uniquely by the sign of the form $f(\eta)$
which can be established only by thermodynamics. In this connection, we note
that functional (\ref{LDE2}) represent in a general form a
Glansdorff-Prigogine local potential. Thus. if the extended MREP principle
establishes the minimum of $\widetilde{\Psi}[\rho]$ , automatically, it also
establishes the maximum of $\Psi\lbrack\rho,\mathbf{J}]$\ along the fixed
force path, thus allowing for an extended LDE principle.

Let us consider the simplest case of a functional dependent on a single state
variable. In the domain $\left\{  \mathbf{X},\mathbf{J}\right\}  $ we have
$\psi=\mathbf{J\cdot X}-(1/2)M_{0}\mathbf{J}^{2}$ where $M_{0}>0$ because of
the second thermodynamic law. The function $\psi$ describes a saddle-shaped
surface where the saddle point $\left\{  X_{0},J_{0}\right\}  $ is a maximum
along the path $\delta\mathbf{X}=0$ and a minimum along the path
$\delta\mathbf{J}=M_{0}\delta\mathbf{X}$. In the important case of the heat
conduction \ we have $X=\nabla(1/T)$ and $L=\lambda(T)T^{2}$ where
$\lambda(T)$ is a temperature-dependent coefficient of thermal conductivity.
The restricted functional is
\[
\widetilde{\Psi}\left[  T\right]  =\frac{1}{2}\int\lambda(T_{0})T_{0}%
^{2}\left[  \nabla(1/T)\right]  ^{2}d^{3}x
\]
which is the local potential for heat conduction$^{8}$.
\[
\]
\textbf{IV} \textbf{SOME EXAMPLES}%
\[
\]
\textbf{Bulk thermoelectricity in intrinsic semiconductors. }We apply the
above variational procedure to the case of bulk thermolectricity in
semiconductors where cross effects between electron-hole and heat transport
are to be taken into account. The entropy production is (compare with eq.
\ref{MREP1} where $-\xi$ is replaced by $\upsilon$)%

\begin{equation}
\sigma_{S}=-\mathbf{j}_{n}\cdot\nabla\frac{\varphi_{n}}{T}-\mathbf{j}_{p}%
\cdot\nabla\frac{\varphi_{p}}{T}+\mathbf{J}_{Q}\cdot\nabla\frac{1}{T}%
+\frac{\upsilon\Gamma}{T}\label{var16}%
\end{equation}
where $\varphi_{n}$ and $\varphi_{p}$ are the electrochemical potentials of
electrons and holes, respectively, ($\varphi_{n}=\mu_{n}+eV$ and $\varphi
_{p}=\mu_{p}-eV$, $V$ standing for electric potential, $e$ for the absolute
value of electron charge and $\mu$ for chemical potentials), $\upsilon$ for
the generation-recombination rate ($\upsilon>0$ for recombinations) and
$\Gamma=\varphi_{n}+\varphi_{p}$ for the carrier affinity (or, equivalently,
for the differences of quasi-Fermi levels$^{4}$). On these grounds, by taking
into account of eq. (\ref{MREP15BIS}) \ (with fluxes as variables) we have%

\begin{equation}
\Psi(\varphi,T,J,\upsilon)=-\int\left(
\begin{array}
[c]{c}%
\frac{1}{2}M_{0nn}~j_{n}^{2}+\frac{1}{2}M_{0pp}~j_{p}^{2}+\frac{1}{2}%
M_{0QQ}J_{Q}^{2}+M_{0nQ}~\mathbf{j}_{n}\cdot\mathbf{J}_{Q}+\\
+M_{0pQ}~\mathbf{j}_{p}\cdot\mathbf{J}_{Q}+M_{0np}\text{~}\mathbf{j}_{n}%
\cdot\mathbf{j}_{p}+\frac{1}{2}M_{0vv}\upsilon^{2}%
\end{array}
\right)  d^{3}x \label{Var20}%
\end{equation}%
\[
-\int\left[  \mathbf{j}_{n}\cdot\nabla\frac{\varphi_{n}}{T}+\mathbf{j}%
_{p}\cdot\nabla\frac{\varphi_{p}}{T}-\mathbf{J}_{Q}\cdot\nabla\frac{1}%
{T}-\frac{\upsilon\Gamma}{T}\right]  d^{3}x
\]
By varying potential\ $\Psi$\ with respect to temperature we obtain%

\[
\nabla\cdot\frac{\partial\sigma_{S}}{\partial\nabla T}-\frac{\partial
\sigma_{S}}{\partial T}=0
\]
that is%

\[
\nabla\cdot\frac{\widetilde{\mathbf{J}}_{Q}}{T^{2}}+2\frac{\widetilde
{\mathbf{J}}_{Q}}{T^{3}}\cdot\nabla T+\mathbf{j}_{n}\cdot\frac{\nabla
\varphi_{n}}{T^{2}}+\mathbf{j}_{p}\cdot\frac{\nabla\varphi_{p}}{T^{2}}%
-\frac{\upsilon\Gamma}{T^{2}}=0
\]
where $\widetilde{\mathbf{J}}_{Q}=\mathbf{J}_{Q}-\sum_{i}\varphi
_{i}~\mathbf{j}_{i}$ so that (see eqs \ref{MREPYY4}-\ref{MREP1} )%

\[
\nabla\cdot\mathbf{J}_{S}=\sigma_{S}%
\]
As for the flux balance equations (variations with respect to $\varphi$), it
is easy to verify that $\nabla\cdot\mathbf{j}_{n}=-\upsilon$ and $\nabla
\cdot\mathbf{j}_{p}=-\upsilon$

Now, let us consider the flux laws (variations with respect to fluxes). As for
electron and hole fluxes, it is to be pointed out that we does not expect
direct cross effect. In this connection, later we will use (even if not
explicitly shown) $L_{np}=(M^{-1})_{np}=0$ , that is, $M_{0QQ}M_{0np}%
=M_{0nQ}M_{0pQ}$. The stationary fluxes are%

\begin{equation}
\mathbf{j}_{n}=-\frac{1}{M_{0nn}}\nabla\frac{\varphi_{n}}{T}-\frac{M_{0nQ}%
}{M_{0nn}}\mathbf{J}_{Q}-\frac{M_{0np}}{M_{0nn}}\mathbf{j}_{p} \label{Var21}%
\end{equation}

\begin{equation}
\mathbf{j}_{p}=-\frac{1}{M_{0pp}}\nabla\frac{\varphi_{p}}{T}-\frac{M_{0pQ}%
}{M_{0pp}}\mathbf{J}_{Q}-\frac{M_{0np}}{M_{0pp}}\mathbf{j}_{n} \label{Var22}%
\end{equation}

\begin{equation}
\mathbf{J}_{Q}=-\frac{1}{M_{0QQ}}\frac{1}{T^{2}}\nabla T-\frac{M_{0nQ}%
}{M_{0QQ}}\mathbf{j}_{n}-\frac{M_{0pQ}}{M_{0QQ}}\mathbf{j}_{p} \label{Var23}%
\end{equation}
where the Onsager reciprocity relations will appear when the forces are
explicitly shown. To write the flux equations in a more compact form it is
convenient to define%

\begin{equation}
\frac{e^{2}}{\sigma_{n}}=\left(  1-\frac{M_{0nQ}^{2}}{M_{0nn}M_{0QQ}}\right)
M_{0nn} \label{Var24}%
\end{equation}

\begin{equation}
\frac{e^{2}}{\sigma_{p}}=\left(  1-\frac{M_{0pQ}^{2}}{M_{0pp}M_{0QQ}}\right)
M_{0pp} \label{Var25}%
\end{equation}
Note that $1-M_{0nQ}^{2}/M_{0nn}M_{0QQ}>0$ and $1-M_{0pQ}^{2}/M_{0pp}%
M_{0QQ}>0$ due to positiveness of the quadratic form. Here $\sigma_{n}$ and
$\sigma_{p}$\ stand for the electron and hole conductivities, respectively.
Thus, current densities of electrons and holes are respectively%

\begin{equation}
\frac{\mathbf{I}_{n}}{-e}=\mathbf{j}_{n}=-\frac{\sigma_{n}}{e^{2}}\left(
T~\nabla\frac{\varphi_{n}}{T}-\frac{M_{0nQ}}{M_{0nn}}\frac{1}{T}\nabla
T\right)  \label{Var26}%
\end{equation}

\begin{equation}
\frac{\mathbf{I}_{p}}{e}=\mathbf{j}_{p}=-\frac{\sigma_{p}}{e^{2}}\left(
T~\nabla\frac{\varphi_{p}}{T}-\frac{M_{0pQ}}{M_{0pp}}\frac{1}{T}\nabla
T\right)  \label{Var27}%
\end{equation}
which should be compared with the well known thermoelectric current
equations$^{15}$%

\begin{equation}
\frac{\mathbf{I}_{n}}{-e}=-\frac{\sigma_{n}}{e^{2}}\left(  \nabla\varphi
_{n}+\frac{Q_{n}}{T}\nabla T\right)  \label{Var17}%
\end{equation}

\begin{equation}
\frac{\mathbf{I}_{p}}{e}=-\frac{\sigma_{p}}{e^{2}}\left(  \nabla\varphi
_{p}+\frac{Q_{p}}{T}\nabla T\right)  \label{Var18}%
\end{equation}
where $Q_{n}$ and $Q_{p}$ stand for the excess of kinetic energy with respect
to the mean energy of electrons and a holes, respectively. The following
correspondence are found $Q_{n}=-\left(  M_{0nQ}/M_{0QQ}+\varphi_{n}\right)  $
and $Q_{p}=-\left(  M_{0pQ}/M_{0QQ}+\varphi_{p}\right)  $ .Based on these
results, the heat flux equation can be reduced to the simple form%

\begin{equation}
\mathbf{J}_{Q}=-\lambda\nabla T+\left(  Q_{n}+\varphi_{n}\right)
\mathbf{j}_{n}+\left(  Q_{p}+\varphi_{p}\right)  \mathbf{j}_{p} \label{Var19}%
\end{equation}
where we used $\lambda=1/M_{0QQ}T^{~2}$. Finally, the following relations are found%

\begin{equation}
M_{0nn}=\left(  e^{2}/\sigma_{n}T\right)  +\left(  1/\lambda\right)  \left[
\left(  Q_{n}+\varphi_{n}\right)  /T\right]  ^{2} \label{Var28}%
\end{equation}

\begin{equation}
M_{0pp}=\left(  e^{2}/\sigma_{p}T\right)  +\left(  1/\lambda\right)  \left[
\left(  Q_{p}+\varphi_{p}\right)  /T\right]  ^{2} \label{Var29}%
\end{equation}

\begin{equation}
-M_{0nQ}=\left(  Q_{n}+\varphi_{n}\right)  /\lambda~T^{~2} \label{Var30}%
\end{equation}

\begin{equation}
-M_{0pQ}=\left(  Q_{p}+\varphi_{p}\right)  /\lambda~T^{~2} \label{Var31}%
\end{equation}

\begin{equation}
M_{0np}=\left(  Q_{n}+\varphi_{n}\right)  \left(  Q_{p}+\varphi_{p}\right)
/\lambda~T^{~2} \label{Var32}%
\end{equation}

As for the local rate of electron-hole generation recombination process, we
obtain $\upsilon=M_{0vv}^{-1}\Gamma/T$ or, by putting $M_{0vv}^{-1}%
=L_{0vv}T_{0}$, $\upsilon=L_{0vv}\Gamma$ (at SNES).%

\[
\]
\textbf{External forces}

The thermodynamics of SNES involve non isolated system because of the
necessity of continuosly feeding the stationary processes. The external
parameters fixing the SNES are not included in the theory which only
establishes the relations among physical quantities of the system dealt with.
In several case this could generate difficulties if internal fluxes to be
varied are bound with the external ones. This problem can be figured out by
means of proper additional constraining equations. However, we may also
conveniently re-define the thermodynamic potentials to include explicitly the
external fluxes. To this purpose, we use a different entropy balance equation
in which negative terms are added to the entropy production. These additional
terms account for the action of external forces continuously displacing the
system from equilibrium. This is tantamount to say that we are considering the
rate of the whole system entropy change, which is expected to vanish at SNES
(the system entropy remain constant). The nature of additional terms
discriminate the internal fluxes (which appear in the kinetic term of
potential $\psi$) from the internal forces. In this connection, it is to be
taken into account that derivatives with respect to forces $\mathfrak{X}%
_{\alpha}$ coniugated to external fluxes $\mathfrak{J}_{\alpha}$ should satisfy%

\begin{equation}
\frac{\partial\psi}{\partial\mathfrak{X}_{\alpha}}=-\frac{\mathfrak{J}%
_{\alpha}}{T}\label{Var33}%
\end{equation}
Thus, we will consider as internal fluxes the ones which links to fixed
$\mathfrak{J}_{\alpha}$. Note that presence of external terms does not change
the conclusions about the potential $\psi$ which depends on the internal
properties of the system. We have only a shift of the stationary point in the
$\psi,J,X$ map. Of course, this topic merges with the boundary problem
discussed in section 5.%

\[
\]
\textbf{The Maxwell theorem of circuits}

To explain the above point, let us consider a network of resistances and
voltage sources $\mathfrak{F}_{\alpha}$ (thus a discrete system), for example
ideal batteries.. All the chemical processes taking place inside a battery are
to be considered as external to the circuit (open system). The whole entropy
of battery increases due to spontaneous processes, but it release a negative
entropy $-I_{\alpha}\mathfrak{F}_{\alpha}\mathfrak{/}T$ \ to the circuit due
to the work of charge separation. The current\ $I_{\alpha}$ crossing the
voltage source depend on the circuit features. Thus, all the circuit currents
are to be considered as forces in the potential $\psi$. Instead, the
corresponding affinities are to be considered as fluxes to be linked with
fixed $\mathfrak{F}_{\alpha}$. On this ground, potential $\psi$ should be
written as%

\begin{equation}
\psi=\sum_{i}\frac{I_{i}~\Gamma_{i}}{T}-\sum_{\alpha}\frac{I_{\alpha
}~\mathfrak{F}_{\alpha}}{T}-\sum_{i}\frac{1}{2}M_{0i}\Gamma_{i}^{2}%
\label{Var34}%
\end{equation}
SNES is to be searched for with additional current constraints, that is,
\begin{equation}
\Omega_{\Gamma}=\sum_{i}\frac{I_{i}~\Gamma_{i}}{T}-\sum_{\alpha}%
\frac{I_{\alpha}~\mathfrak{F}_{\alpha}}{T}-\sum_{i}\frac{1}{2}M_{0i}\Gamma
_{i}^{2}+\sum_{\beta}\lambda_{\beta}~f_{\beta}(I_{\left\{  i\right\}
},I_{\left\{  \alpha\right\}  })\label{Var35}%
\end{equation}
Alternatively, if we deal with current sources $\mathfrak{I}_{\alpha}$ (now
the fixed external fluxes), SNES is to be searched for by means of%

\begin{equation}
\psi=\sum_{i}\frac{I_{i}~\Gamma_{i}}{T}-\sum_{\alpha}\frac{\mathfrak{I}%
_{\alpha}~\Gamma_{\alpha}}{T}-\sum_{i}\frac{1}{2}L_{0i}I_{i}^{2} \label{Var41}%
\end{equation}
and

.%
\begin{equation}
\Omega_{I}=\sum_{i}\frac{I_{i}~\Gamma_{i}}{T}-\sum_{\alpha}\frac
{\mathfrak{I}_{\alpha}~\Gamma_{\alpha}}{T}-\sum_{i}\frac{1}{2}L_{0i}I_{i}%
^{2}+\sum_{\beta}\eta_{\beta}~g_{\beta}(\Gamma_{\left\{  i\right\}  }%
,\Gamma_{\left\{  \alpha\right\}  }) \label{Var36}%
\end{equation}
where $g_{\beta}(\Gamma_{\left\{  i\right\}  },\Gamma_{\left\{  \alpha
\right\}  })$ are the affinity constraints.

As example, let us consider the simple circuit formed by a battery
$\mathfrak{F}$ and two parallel resistances, $R_{1}$ and $R_{2}$, connected to
a resistance $R_{3}$. Current constraints are $I=I_{1}+I_{2}=I_{3}$. Thus,%
\begin{equation}
\Omega_{\Gamma}=\sum_{i}\frac{I_{i}~\Gamma_{i}}{T}-\frac{I~\mathfrak{F}}%
{T}-\sum_{i}\frac{1}{2}M_{0i}\Gamma_{i}^{2}+\lambda_{a}~(I-I_{2}%
-I_{1})+\lambda_{b}(I-I_{3})\label{Var37}%
\end{equation}
From  $\partial\Omega_{\Gamma}/\partial\Gamma_{i}=0$, $\partial\Omega_{\Gamma
}/\partial I_{i}=0$ and $\partial\Omega_{\Gamma}/\partial I=0$\ \ it follows
$\Gamma_{i}=I_{i}/\left(  T~M_{0i}\right)  $, $\Gamma_{1}=\Gamma_{2}%
=T~\lambda_{a}$, $\Gamma_{3}=T~\lambda_{b}$ and $\mathfrak{F=}\Gamma
_{1}+\Gamma_{3}=\Gamma_{2}+\Gamma_{3}$, respectively. As for potential
$\widetilde{\psi}$ (for variations of $\psi$ along $\delta I_{i}=M_{0i}%
~\delta\Gamma_{i}$ paths) we have%

\begin{equation}
\widetilde{\psi}=\sum_{i}\frac{1}{2}\frac{R_{i}I_{i}^{2}}{T}-\frac
{I~\mathfrak{F}}{T}\label{Var38}%
\end{equation}
where $R_{i}=\left(  T~M_{0i}\right)  ^{-1}$ . Equation (\ref{Var38}) (or its
generalization) by addition of current constraints call for the Maxwell
theorem of circuits$^{16}$.%
\[
\]
\textbf{Photo-excitations in semiconductors and recombination resistances. }

Now let us consider the electron-hole recombinations in semiconductors under
steady injection of carrier in conduction and valence bands. For simplicity,
we will deal with the case of uniform photo-excitation and temperature.
\ According to these settings we are assuming the infinite thermal
conductivity of the lattice to which the heat produced by energy conversion
(which accounts for the external entropy production) is released. Blackbody
field is taken into account in rate equations. Of course, the uniformity
conditions make the foregoing considerations suitable for the system near
equilibrium.\ Potential $\psi$ becomes
\begin{equation}
\psi=\frac{\upsilon\Gamma}{T}-\frac{\Phi\Gamma}{T}-\frac{1}{2}M_{0vv}%
\upsilon^{2}\label{Var39}%
\end{equation}
where $\Phi$ stands for the rate of (externally induced) carrier generation.
The sign minus account for the negative contribution to the entropy induced in
the system by the external force. Really, it is not relevant how the system is
moved from equilibrium since we are dealing with a fixed rate $\Phi$. At the
steady state we obtain the expected equation $\upsilon=L_{0vv}\Gamma_{0}=\Phi
$. As for $\widetilde{\psi}$ (for variations of $\psi$ along $\delta
\upsilon=L_{0vv}\delta\Gamma$ path) it holds%

\begin{equation}
\widetilde{\psi}=\frac{1}{2}\frac{L_{0vv}\Gamma^{2}}{T}-\frac{\Phi\Gamma}%
{T}\label{Var40}%
\end{equation}
Finally, it is easily recognizable that eq. (\ref{Var39}) is similar to
eq.(\ref{Var41}). Thus we can conclude that at SNES (near equilibrium) a
Maxwell theorem can be called for the recombination rate distribution in a
semiconductor analogously to the distribution of currents in a electric
ciurcuit; now resistances are to be replaced by recombination resistances
related to the inverse of recombination probabilities$^{4}$%

\[
\]
\textbf{V.} \textbf{FINAL REMARKS AND CONCLUSIONS.}

We still remark that the variational equation (\ref{MREPZZ15})\ stands on the
condition of having fixed state variables at the system boundary. We can
remove this condition by considering a surface integral such that$^{8}$.%

\begin{equation}
\int\sum_{k}\mathbf{J}_{0k}\cdot\delta\mathbf{X}_{k}dV-%
%TCIMACRO{\doint _{Surf}}%
%BeginExpansion
{\displaystyle\oint_{Surf}}
%EndExpansion
\sum_{k}\left[  \delta\left(  \frac{\varphi_{k}}{T}\right)  \right]
\mathbf{J}_{0k}\cdot d\mathbf{A}=0 \label{MREPXY1}%
\end{equation}
Consequently eq. (\ref{MREPZZ15}) is to be replaced by%

\begin{equation}
\delta\left(  \Psi-%
%TCIMACRO{\doint _{Surf}}%
%BeginExpansion
{\displaystyle\oint_{Surf}}
%EndExpansion
\sum_{k}\frac{\varphi_{k}}{T}\mathbf{J}_{0k}\cdot d\mathbf{A}\right)  =0
\label{MREPXY2}%
\end{equation}
We need not modify anything in the conclusion of the previous section
concerning the extremum properties of the functional $\Psi$. Indeed, eq.
(\ref{MREPXY1}) holds at any variation order \ so that the second variation of
the modified functional coincides with $\delta^{(2)}\Psi$.

In the case $grad~\mathbf{v}\neq0$ we must include the viscoelastic term in
the entropy production.$^{5,12}$ In this connection we point out that
$\mathbf{v}$\ cannot be considered as a functional variable. This is because
the variational criterion (\ref{MREPZZ15})\ is closely related to the entropy
conservation law and does not account for momentum conservation. Thus, the
functional (\ref{LDE2}), as it stands, cannot be used in the field of
hydrodynamics. We need not go further in this matter which is beyond the scope
of this paper.

We have shown that in continuous systems satisfying LTE conditions, the
(extended) MREP and (local) LDE principles are connected by means of a
suitable functional. In the case of a single state variable, the density of
the functional describes a saddle-shaped surface in the force-flux domain. The
saddle point represents the stationary state. The MREP path is the one where
the phenomenological coefficient maintains its reference state (SNES) value.
Instead, the LDE path is the one where the force is kept fixed. In particular,
along the MREP path the functional behaves as a Glansdorff-Prigogine local potential.%

\[
\]

$^{a}$E-mail:masalis@unica.it

$^{1}$. I$.$Prigogine, Introduction to Thermodynamics of Irreversible
Processes , (John Wiley \& Sons, Interscience, N. Y., 1971); G. Nicolis and I.
Prigogine, Self-Organization in Nonequilibrium Systems (John Wiley \& Sons New
York, London, Sydney, Toronto, 1977).

$^{2}$. D. K. Kondepudi, Physica A \textbf{154}, 204 (1988);K. L. C. Hunt , P.
M. Hunt and J. Ross, Physica A \textbf{154}, 207 (1988).

$^{3}$. P. H. E. Meeijer and J. C. Edwards Phys. Rev A \textbf{1}, 1527(1970).

$^{4}$. M. Salis, P. C. Ricci and F. Raga, Phys. Rev. B \textbf{72}, 35206.(2005).

$^{5}$. H. J. Kreuzer, Nonequilibrium Thermodynamics and its Statistical
Foundations (Clarendon Press Oxford,1981).

$^{6}$. D. D. Fitts, Nonequilibrium Thermodynamics (McGraw-Hill Book Company,
inc. New York 1962).

$^{7}$. P. Palffy-Mhuroray, Am. J. Phys, \textbf{69}, 825 (2001); W. G.
Hoover, Am. J. Phys. \textbf{70}, 452 (2002).

$^{8}$. P. Glansdorff and I. Prigogine, Physica \textbf{30}, 351 (1964).

$^{9}$. P. Van and W. Mushik, Phys. Rev E \textbf{52}, 3584 (1995).

$^{10}$. L. Onsager, Phys. Rev., \textbf{37},405 (1931); \textbf{38},2265 (1931).

$^{11}$. L. Onsager and S. Machlup, Phys. Rev \textbf{91}, 1505 (1953); S.
Machlup and L. Onsager, Phys. Rev \textbf{91}, 1512 (1953).

$^{12}$. S. R. De Groot, Non-Equilibrium Thermodynamics, Rendiconti della
Scuola Internazionale di Fisica "E. Fermi", Ed. S. R. De Groot (Societ\`{a}
Italiana di Fisica, Nicola Zanichelli Editore, Bologna 1960).

$^{13}$. W. Yougrau, A. Van der Merwe and Gough Raw, Treatise on Irreversible
and Statistical Thermodynamics (Macmillan N.Y. Collier-Macmillan London 1966).

$^{14}$. C. A. Domenicali, Rev. Mod. Phys. \textbf{26}, 237 (1954).

$^{15}$ J. Tauc, Rev. Mod. Phys. \textbf{29}, 308 (1957).

$^{16}$ J. P. P\`{e}rez Am. J. Phys. \textbf{68}, 860 (2000)%

\[
\]

\end{document}